\documentclass[manuscript]{emulateapj}
\usepackage{epsf,wrapfig}
\usepackage{graphicx}
\usepackage{amsmath}

\begin{document}
  
\title{The Pulsar B2224+65 and its Jets: A Two Epoch X-ray Analysis}
\author{S. P. Johnson \& Q. D. Wang}
\affil{Department of Astronomy, University of Massachusetts Amherst,
  Amherst, MA}

\begin{abstract}
We present an X-ray morphological and spectroscopic study of the
pulsar B2224+65 and its apparent jet-like X-ray features based on two
epoch \textit{Chandra} observations. The main X-ray feature, which 
shows a large directional offset from the ram-pressure confined pulsar
wind nebula (Guitar Nebula), is broader in apparent width and shows
evidence for spectral hardening (at 95 percent confidence) in the
second epoch compared to the first.  Furthermore, the sharp leading edge of
the feature is found to have a proper motion consistent with that of
the pulsar ($\sim$180 mas yr$^{-1}$).  The combined data set also
provides evidence for the presence of a counter feature, albeit
substantially fainter and shorter than the main one.  Additional
spectral trends along the major and minor axes of the feature are only
marginally detected in the two epoch data, including softening counter
to the direction of proper motion.  Possible explanations for the
X-ray features include diffuse energetic particles being confined by an
organized ambient magnetic field as well as a simple ballistic jet
interpretation; however, the former may have difficulty in explaining
observed spectral trends between epochs and along the feature's major
axis whereas the latter may struggle to elucidate its linearity.
Given the low counting statistics available in the two epoch
observations, it remains difficult to determine a physical production
scenario for these enigmatic X-ray emitting features with any certainty. 
\end{abstract}

\section{Introduction}

As pulsars age, they lose their spin-down energy through both
radiation and relativistic particle ejection. One may approximate the
particle ejection in two forms: anisotropic winds from the pulsar
surface and bi-polar jets.  Previous X-ray studies have shown the wind
component to be primarily equatorial and axisymmetric, producing
observed tori in pulsar wind nebulae (PWNe), whereas the jet outflows
are only observed in a few, though well studied, young pulsars
\citep[e.g. the Crab and Vela pulsars,][]{weiss00,pavlov03}.  Only a
few nearby and relatively old pulsars have been shown to have their
PWNe, mostly enhanced by ram-pressure confinement \citep[e.g.,][and
  references therein]{wang93,karga08}. For a PWN or similar pulsar ejection
feature to be observed in X-rays, the synchrotron-emitting particles
must have energies on order of 10$^2$ TeV, achievable in
shocks of the pulsar ejecta as it encounters the local interstellar
medium (ISM). X-ray investigations of pulsar outflows are therefore
crucial in understanding their origin and particle acceleration, which
can in turn provide insight into the origin of other, more exotic and
distant, outflows (e.g. from Active Galactic Nuclei).

A particularly striking case is that of the pulsar B2224+65 and its
X-ray-emitting ``jet''.  B2224+65 is one of the fastest pulsars known
with a radio proper motion of $\mu=182\pm 3$ mas yr$^{-1}$
\citep{harrison93}; corresponding to a transverse
velocity of $864 \rm~D_{kpc}~km~s^{-1}$, where $\rm~D_{kpc}$ is the
distance to B2224+65 in units of kpc and is uncertain in the range of
$\sim 1-2$ \citep{chat04}.  The rapid motion of the pulsar produces a
bow shock nebula visible in H$\alpha$ \citep{cordes93,chat02,chat04}
which has been dubbed the ``Guitar Nebula'' due to its peculiar
morphology.  Aside from its high proper motion, B2224+65 behaves like
a normal radio pulsar, with a modest energy spin-down rate of
$\dot{E} = 10^{33.1}\rm~ergs~s^{-1}$ and period $P=0.68$ s.  X-ray
studies of B2224+65 do not reveal significant emission concurrent with
the H$\alpha$ \citep[however, see][]{romani97} but rather an unusual
extended linear feature apparently stemming from B2224+65 and offset
from the pulsar's direction of motion by $\sim$118$\degr$
\citep{hui07,wong03,zavlin04}.  The feature, extending $\sim$2$\arcmin$
away from the pulsar, was found to have an apparent non-thermal
spectrum though the details as to its origin have thus far remained a
mystery. Such a distinct, linear X-ray emitting feature has not
been identified anywhere else in the Galaxy with the possible 
exception of the Galactic Center \citep[e.g.][and references
  therein]{johnson09}.  It was suggested by \cite{karga08} that the
extended feature may not be associated with B2224+65 but rather with a
nearby point source, although no physical production scenario was
developed for the feature. Through theoretical modeling of the linear
X-ray emission by \cite{band08}, it was proposed that the feature
could be produced through highly energetic particles that are
accelerated at, and subsequently escape, the pulsar wind terminal shock but are
ultimately confined by a large-scale, organized interstellar magnetic
field.  However, there are a number of observational requirements to
the \cite{band08} model that have yet to be fulfilled, primarily a
proper motion of the feature consistent with B2224+65.  The uncertain
association with the pulsar, combined with low photon statistics, has
made it difficult to establish the physical process responsible for
the feature. 

\begin{deluxetable*}{lcccccc}
\tablecaption{\textit{Chandra} ACIS-S Observations}
\tablehead{\textit{Chandra} &R.A. (J2000) &Dec. (J2000) & Exposure &Roll Angle &OBS Date\\
OBS. \#  &(h~~m~~s) &(degree~~arcmin~~arcsec)&(s) &(degree) &(yyyy-mm-dd)}
\startdata
755      &22 25 48.439 & +65 35 08.09 &  47865 &   236.8 &2000-10-21 \\
6691     &22 25 51.852 & +65 35 39.19 &   9968 &   182.7 &2006-08-29 \\
7400     &22 25 51.428 & +65 35 35.13 &  35839 &   219.3 &2006-10-06
\enddata
\tablecomments{The exposure represents the live time (dead time
  corrected) of cleaned data.}
\end{deluxetable*}

\begin{figure*}
\includegraphics{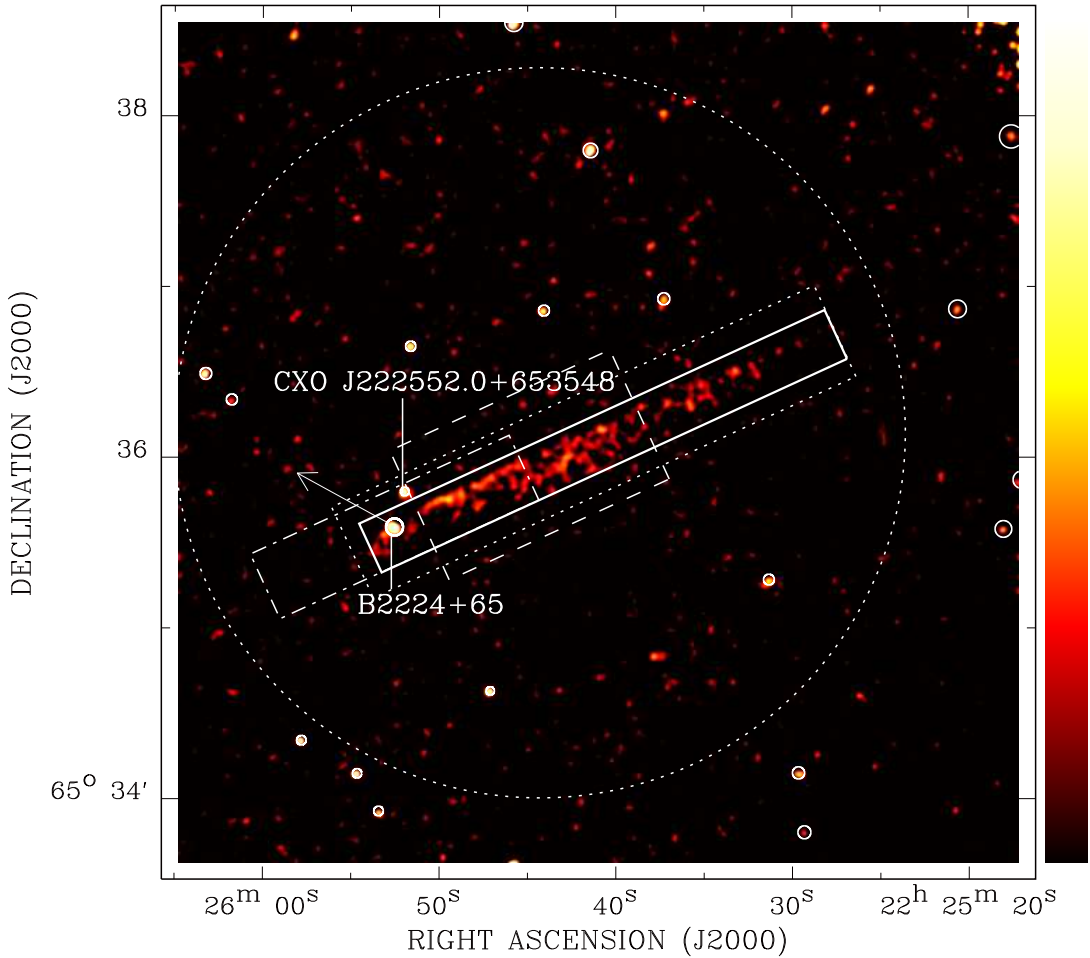}
\caption{Merged epoch \textit{Chandra} intensity map of B2224+65 in
  the 0.3-7.0 keV energy band.  The image has been smoothed using an
  adaptive Gaussian filter (FWHM$\la$1$\arcsec$) and is presented with
  logarithmic scaling.  Overlaid on the intensity map is the direction
  of B2224+65's proper motion along with detected X-ray sources.  Also
  included are the regions used for spectral extraction of the pulsar and
  jet (solid line), the local background region (the area within the
  dotted line excluding sources and the dotted rectangle), and the
  regions used in constructing surface brightness profiles along the
  major (dot-dash line) and minor (dashed line) axis.  These regions
  all have position angles of $\sim$24.7$\degr$ as determined from the
  bright edge of the jet within $\sim$50$\arcsec$ from B2224+65.}
\end{figure*}

In this study, we examine two epoch data taken by the \textit{Chandra}
X-ray Observatory to determine any association between the extended,
linear feature and B2224+65 as well as to place constraints on possible
X-ray-emitting particle production scenarios.  Throughout this
manuscript, we will refer to the feature as the jet or jet-like
feature based purely on its apparent morphology, implying no physical
interpretation given its still uncertain origin.  From the
two epoch data, we first correct for any astrometric offset between
observations and verify the motion of B2224+65.  We then briefly
examine the two epoch and combined spectroscopic information for the
pulsar to see if there may have been any evolution in its spectrum.  A
morphological analysis of the jet will serve to detect any shift
between epochs, providing evidence for (or against) an association
between the jet and pulsar.  Spectral analysis of the jet between
epochs and across the major/minor axes may aid in identifying possible
spectral trends, which can assist in determining the physical models
behind the X-ray emission. Finally, we briefly discuss our findings in
the context of the \cite{band08} model and a simple ballistic jet
model similar to that used to explain AGN jet emission.

\begin{figure*}[t]
\includegraphics[width=\linewidth]{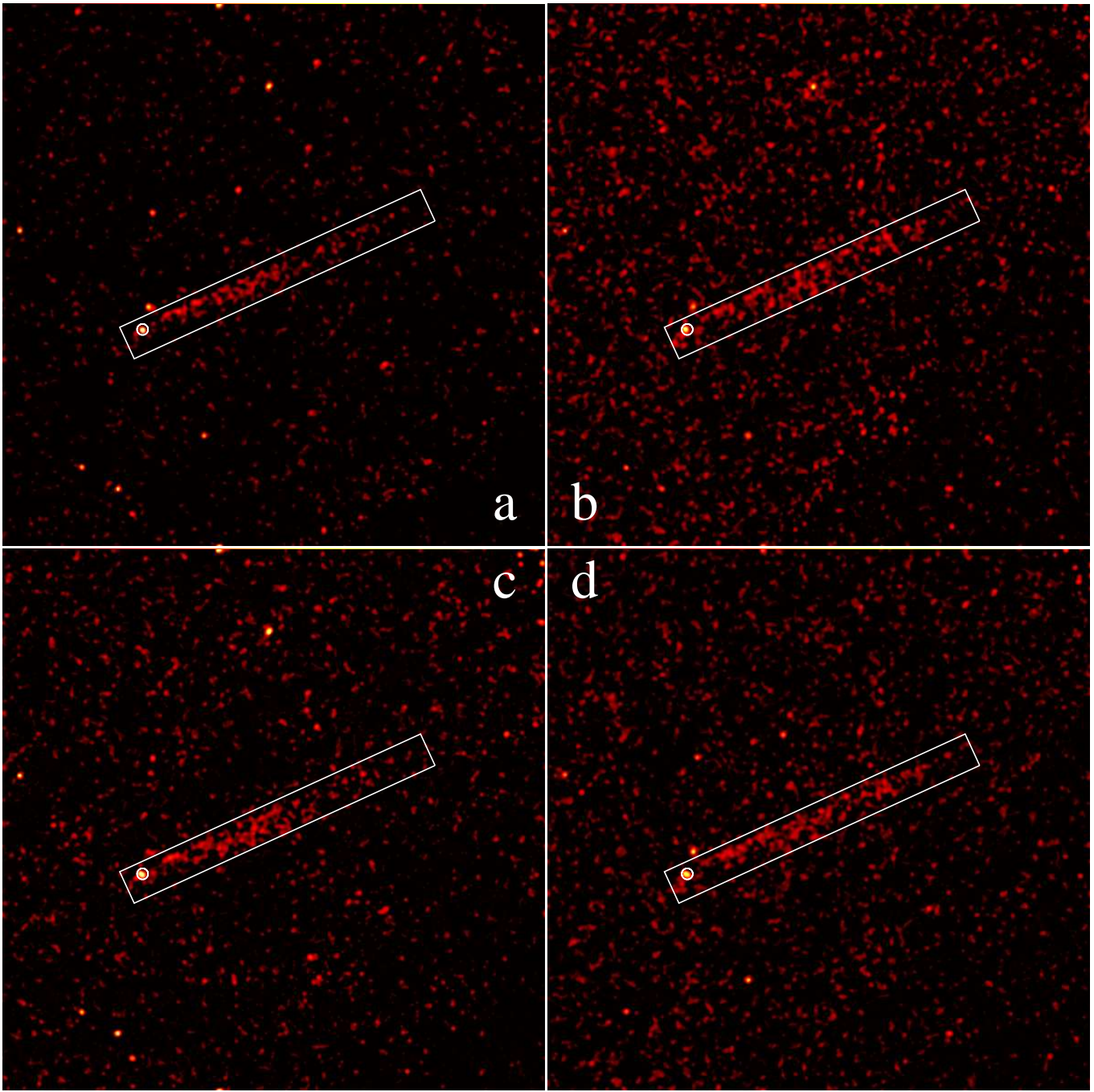}
\caption{Intensity maps of B2224+65 and the jet with the
  source region from Fig.~1 overlaid: 0.3-7.0 keV images in the
  first (a) and second (b) epochs as well as the merged images in the soft
  (0.3-1.5 keV; c) and hard (1.5-7.0 keV; d) bands.  All images are
  adaptively smoothed similar to Fig.~1 and
  are shown here using the same relative logarithmic scale with the
  minimum set by the median image value and the maximum set a factor
  of 10$^3$ times higher.}
\end{figure*}

\section{Observations and Data Reduction}

B2224+65 was observed by the \textit{Chandra} X-ray Observatory over
two separate epochs using the Advanced CCD Imaging Spectrometer
spectroscopic array (ACIS-S).  The first epoch consists of Obs.~ID 755
while the second epoch is composed of two observations, Obs.~ID 6691
and 7400 (the relevant observational parameters for each Obs.~ID are
given in Table 1).  Each of the individual observations was first
reprocessed using the standard \textit{Chandra} Interactive Analysis
of Observations (\textsc{ciao}) routines (version 3.4.0).  After
reprocessing, Obs.~ID 6691 and 7400 were combined
to produce a second epoch event file.  A merged epoch
event file and X-ray image of B2224+65 (Figs.~1,2) was then produced by
combining Obs.~ID 755 and the second epoch event file after
astrometric corrections have been applied (see below).  Source lists
for each of the two epoch event files were constructed using the method
of \cite{wang04}, with a false detection probability of 10$^{-6}$, to be
used in correcting the astrometry between epochs and producing
source-removed event maps for background estimation.

Before we can properly analyze the two epoch data sets, we must
correct for any possible astrometric offset that may have been
produced through pointing differences between epochs.  From our
X-ray detected source lists, we extract the positions of 19 sources
common to both epochs with count rates $\ga$0.2 counts ks$^{-1}$
and off-axis distances $<$4\arcmin.  Limiting the off-axis distance
of the sources serves to minimize the effects of positional errors due
to PSF variation across the CCD.  The RA and DEC offsets, as
  well as a rotation angle in image alignment, are
then derived from the RA and DEC differences of the common sources,
taking into account their positional uncertainty determined from
maximum likelihood centroiding (with 1$\sigma$ errors of
$\sim$0.1-0.9\arcsec).  From our initial calculation, we find that the
rotation between images is negligible (0.02$\pm$0.05\degr,
1$\sigma$ uncertainty) and subsequently recompute the corrections
assuming no rotation, resulting in $\Delta$RA and $\Delta$DEC
corrections of 0.24$\pm$0.09$\arcsec$ and 0.18$\pm$0.09$\arcsec$,
respectively.  Despite exclusion of the rotation in image alignment,
it is unlikely to influence our results, as we limit our analysis to
only the area near the telescope aim-point.  These astrometric
corrections have been applied only to the astrometry in the fits
header of the second epoch event file for merged epoch image
production (Figs.~1,2) and subsequent analyses.

Due to the diffuse nature of the jet, we adopt the method of local
background estimation to determine the background spectra.  To obtain
a representative background for the pulsar and jet regions, we extract
the spectrum from a $\sim$2$\arcmin$ circular region, given by the
dotted circle in Fig.~1, while excluding detected point sources and
the jet.  Since there may be additional, unresolved emission
from the jet extending beyond its source region, we exclude the area
within the dotted rectangle of Fig.~1 when extracting the
background.  This same area is also used for determining the sky background
level when constructing surface brightness profiles along the major
and minor axes of the jet (\S~3.2).

\section{Analysis and Results}

\subsection{Pulsar B2224+65: Proper Motion and Spectrum}

Due to the time gap between the observations of B2224+65, we must
first verify the pulsar's proper motion \citep[$\mu_{RA}=144\pm3$ and
  $\mu_{DEC}=112\pm3$ or $\mu=182\pm3$ mas yr$^{-1}$ with 90 percent
  confidence intervals from the radio sample of][]{harrison93}.  Extracting
B2224+65's position from our X-ray source lists in both epochs and
applying the astrometric corrections, we find that the pulsar's
position has changed by 0.91$\pm$0.24 arcsec in RA and 0.44$\pm$0.24
arcsec in DEC, taking into consideration the relative astrometric
uncertainties, or by 1.01$\pm$0.34 arcsec over the $\sim$6 year gap
between epochs. This relates to a proper motion of
$\mu_{RA}\approx 153\pm 40\rm~mas~yr^{-1}$ and $\mu_{DEC}\approx 74\pm
40\rm~mas~yr^{-1}$ or absolute proper motion of $\mu\approx 170\pm
57\rm~mas~yr^{-1}$ (1$\sigma$ errors), in agreement with the radio
proper motion. 

\begin{deluxetable*}{lccccr}
\tablecaption{Spectral Fitting of B2224+65}
\tablehead{Epoch& Count Rate (0.3-7.0 keV)& N$_H$ & $\Gamma$ & Flux
  (0.3-7.0 keV) &$\chi^2/d.o.f.$\\
& 10$^{-3}$ cts/s& 10$^{22}$ cm$^{-2}$&&10$^{-14}$ ergs cm$^{-2} s^{-1}$}
\startdata
1     & 1.73$\pm$0.19& $<$0.09&  1.89$^{+0.84}_{-0.50}$& $1.71^{+0.44}_{-0.64}$ & 8.94/12\\
2     & 2.00$\pm$0.21& \nodata&  1.66$^{+0.50}_{-0.26}$& $1.91^{+0.50}_{-0.62}$ &\nodata\\
Merged& 1.86$\pm$0.14& $<$0.09&  1.70$^{+0.46}_{-0.23}$& $1.83^{+0.22}_{-0.36}$ & 9.47/14
\enddata
\tablecomments{\textsc{xspec} spectral fitting parameters for
  B2224+65.  The column densities listed are the 90 percent confidence
  upper limits given by the absorbed power law fits. Errors are given
  to 90 percent confidence.} 
\end{deluxetable*}

While we do not expect variation in the spectrum of B2224+65 itself,
a significant portion of the observed emission may arise from the
shocked wind material, which is dependent on the pulsar wind and local
ISM properties, and could therefore change between epochs.  Any
variation, or lack thereof, in the emission can then be compared
with the spectral modeling of the jet.  To ensure that the
pulsar's proper motion will not influence the spectral modeling, we
extract the spectra from the individual epoch event files using a small
circular region (radius of 3$\arcsec$ at RA=22h25m53s,
DEC=+65\degr35\arcmin35$\arcsec$ (J2000), see Fig.~1).
The spectra are then adaptively binned such that each bin has
a signal-to-noise (S/N) ratio of $\ge$2.5 where the signal is the
net counts above the local background and the noise contains all
counts.  This provides 8(9) bins for the first(second) epoch
spectrum with $\sim$10 counts per bin due to the small contribution
from the local background over the pulsar region.

To model the two epoch pulsar spectra, we apply a joint fit, absorbed
power law in \textsc{xspec} (version 11.3.1) where the foreground
X-ray-absorbing column density, N$_{\rm{H}}$, should not change
significantly with time and is thus set as a common parameter between 
epochs.  The epoch averaged properties of the pulsar are derived by
performing a similar joint fit in which all model parameters are set to be the
same in both epochs (i.e. power law index and normalization).
Alternatively, one may also combine the two epoch spectra using the
\textsc{addspec} routine in \textsc{ftools} and apply the absorbed
power law model, or directly extract the spectrum from the merged event
file.  However, the latter method does not preserve calibration
information from each epoch whereas the former may give misleading
results if the detector response changes between spectra and/or if
there is any spatial variation in the spectra.  Nevertheless, we find
that each method produces consistent results with the simple two epoch joint
fit.  We report here only the spectral parameters as given by the
joint fit modeling as it is likely the least sensitive to any
systematic calibration uncertainties. Comparison of our spectral
fitting results (with 90 percent confidence contours given in Table 2
and Fig.~3) of B2224+65 with the previous work of \cite{hui07}
(1$\sigma$ errors based on first epoch data alone) suggests that the
observed emission from the pulsar has not had any significant
evolution between epochs. 

\begin{figure}
\includegraphics[width=.5\textwidth]{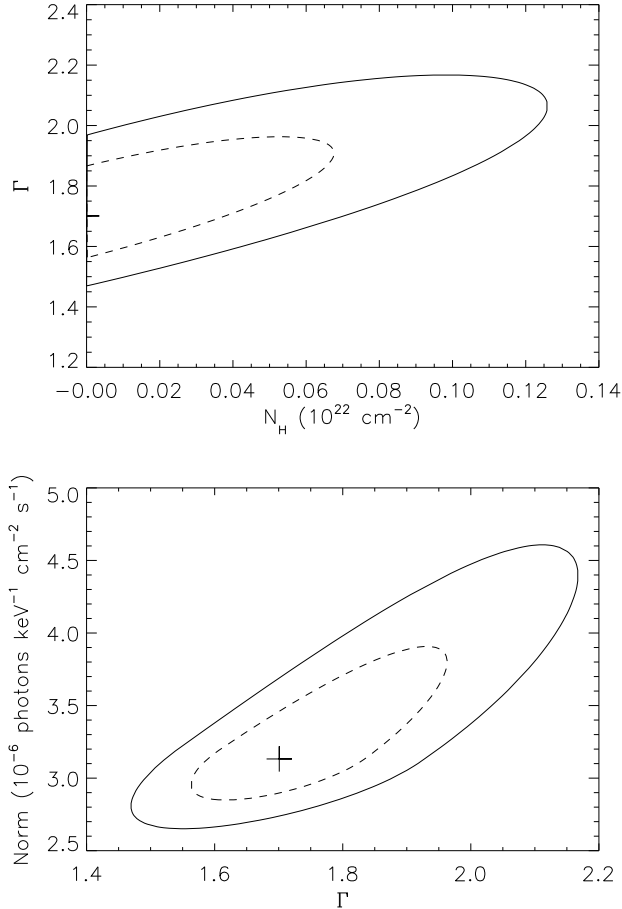}
\caption{68 percent (dashed line) and 90 percent (solid line )
  confidence contours for the epoch averaged pulsar spectral fit for
  N$_{\rm{H}}$ versus $\Gamma$ and for $\Gamma$ versus normalization.
  The crosses in the plots mark the best-fit parameter values from the
  joint fit.}
\end{figure}

\subsection{Jet-Like Feature: Astrometry and Spectrum}

From the smoothed X-ray intensity map of B2224+65 (Fig.~1), the jet
appears disconnected from both B2224+65 and the nearby point source
(labeled here as CXO J222552.0+653548) and thus does not clearly
support a direct association with either source (contrary to the
suggestion of \citealt{karga08}).  The jet remains fairly linear
with a sharp edge along the direction of B2224+65's proper motion,
extending a length of $\sim$2.4\arcmin away from the pulsar. Other
morphological properties include apparent ``fanning'' with increasing distance
from the pulsar and a minor region of X-ray enhancement on the
opposite side of the pulsar.  Furthermore, the two epoch images
(Fig.~2a,b) indicate an increase in the width of the jet across
epochs with little to no bending in the leading edge
($\lesssim$1$\degr$ within $\sim$2$\arcmin$ away from B2224+65) while
the soft and hard band images (Fig.~2c,d) show evidence for harder
X-ray emission farther along the (epoch averaged) jet. In the
following analysis, we assume the case of no bending for simplicity.
The presence of any bending will serve primarily to dilute the
significance of any determined edge location along the minor axis and
will be propagated through our analysis as a systematic uncertainty.

For our morphological analysis of the jet, particularly for
studying its proper motion and the significance of the extended
emission regions, we construct surface brightness profiles along its
major and minor axes. These profiles result from 100$\times$25
arcsec$^2$ and 85$\times$50 arcsec$^2$ rectangular regions centered at
RA=(22h25m52s, 22h25m45s) and DEC=(+65\degr35\arcmin35\arcsec,
+65\degr35\arcmin57\arcsec) (J2000) for the major and minor axis,
respectively (given as the dot-dash and dashed line regions in
Fig.~1).  These regions were selected by hand to maximize the signal
for the relevant profile using the epoch averaged pulsar position as
the primary reference point (\S~3.1). For the calculation of
the profiles, we bin the profiles with a S/N ratio of $\ge$4; here the
signal is the net counts above the instrument background, as
measured from the ACIS stowed data, while the noise includes all
counts.  The use of the instrument background, instead of the
total sky background, enables the application of the adaptive binning
even to regions with negligible contribution from the jet.  For each bin, the
X-ray intensity is then simply defined as (counts-instrument
background)/exposure.  After binning, the sky background levels,
determined from the local background region, were subtracted to
produce the surface brightness profiles shown in Fig.~4.  The adopted
binning was chosen simply for ease of viewing; the exact choice does
not influence our analysis of the jet edge or extended emission
regions (see below).

\begin{figure}
\includegraphics[width=.5\textwidth]{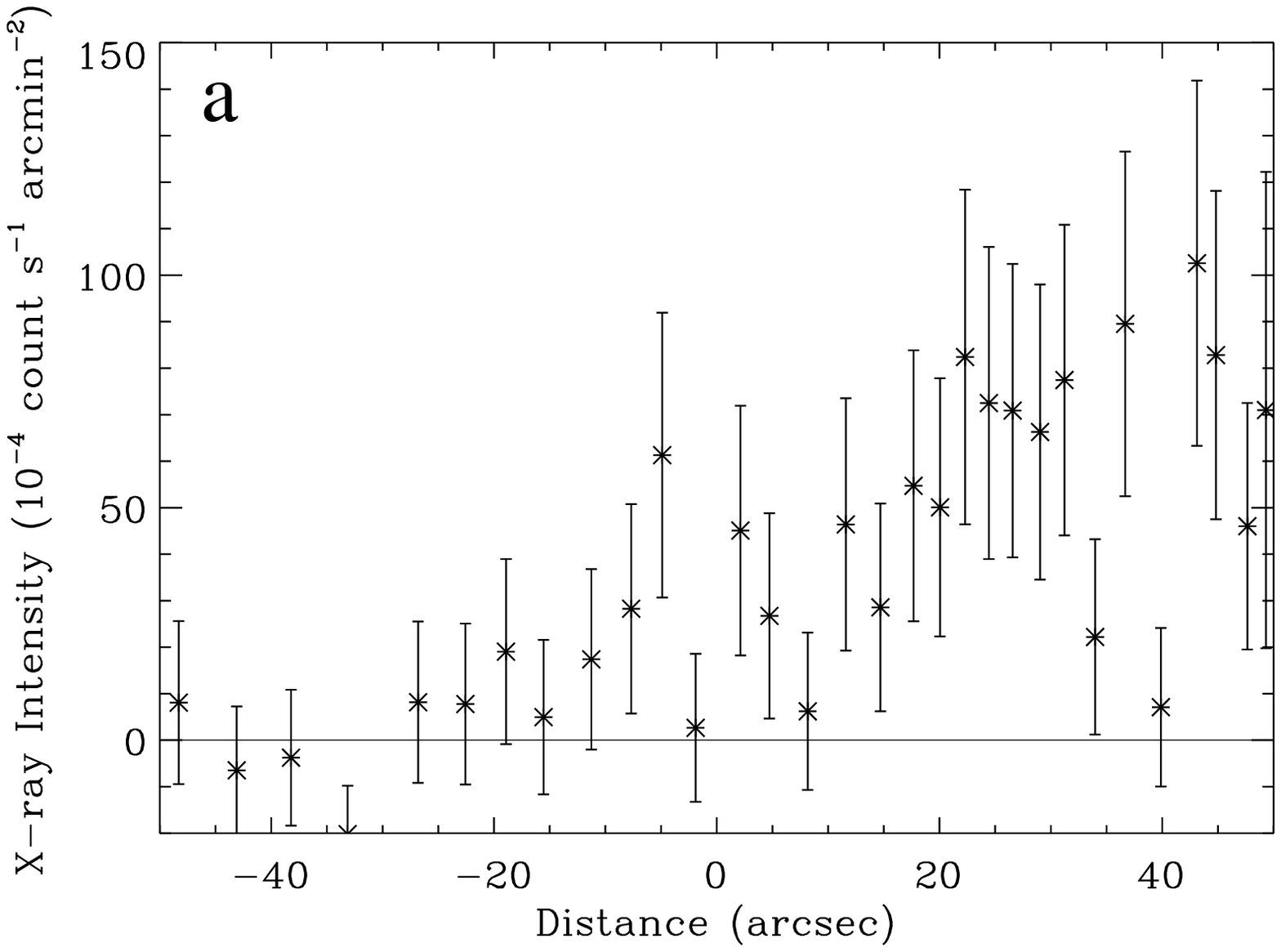}
\includegraphics[width=.5\textwidth]{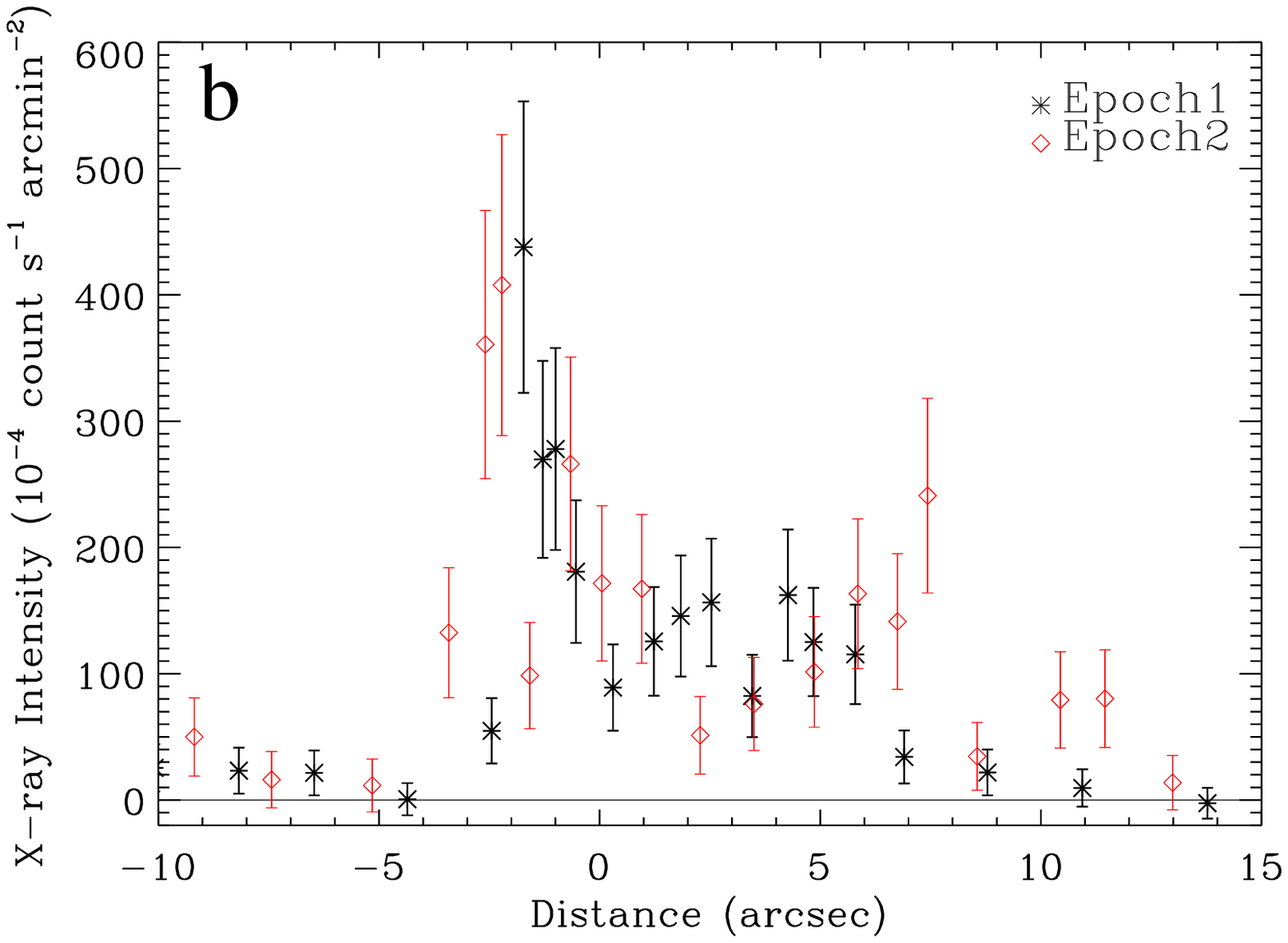}
\caption{X-ray intensity above the sky level along the
 major (a) and minor (b) axes, adaptively binned to a
 S/N of $\ga$4 per bin.  For reference, the sky background
 levels for each profile are 51.9$\pm$0.9, 74.5$\pm$1.2, and
 62.6$\pm$0.7 $\times$10$^{-4}$ counts s$^{-1}$ arcmin$^{-2}$ for
 the first epoch minor axis, second epoch minor axis and combined
 epoch major axis profiles, respectively.  Distances are calculated relative
 to the epoch averaged pulsar position.  Errors are presented at the
 1$\sigma$ confidence level.}
\end{figure}

Examining Fig.~4a, there is evidence for a weak counter feature,
opposite in direction to the primary jet-like feature.  This counter
feature, extending $\sim$20$\arcsec$ away from B2224+65, is
considerably fainter than the main jet-like feature though it still appears
significant compared to the local X-ray background.  Due to the
faintness of the counter feature, the counting statistics are too low
in individual epoch observations for a positive detection. We
therefore use the merged epoch event file, along with a small
20$\times$20 arcsec$^2$ box centered on the feature in question, to
derive a background subtracted count rate of $0.34\pm
0.10\rm~counts~ks^{-1}$.  Comparatively, the local background
corresponding to the same region has a count rate of
0.68$\pm$0.09$\rm~counts~ks^{-1}$, placing the counter feature at 4
times the background uncertainty or at a detection significance of
$\sim$4$\sigma$.  It is therefore highly unlikely that this
feature results from Poisson variations in the X-ray background.  

\begin{figure}
\includegraphics[width=0.5\textwidth]{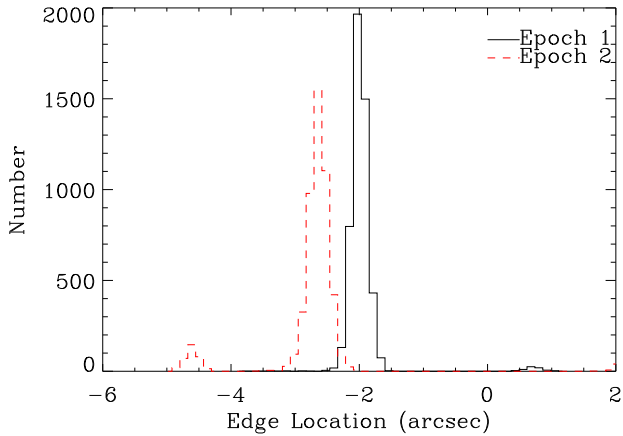}
\caption{Histogram distributions of edge locations in the first and
  second epochs.  The distributions result from bootstrap resampling
  of count profiles along the minor axis, taken from a fine-binned
  count map, and convolving the realized profiles with the chosen
  filter.  The x-axis gives edge locations relative to the epoch
  averaged pulsar position in arcsec.}
\end{figure}

In Fig.~4b, there appears to be a slight shift in the sharp left edge
as well as an extension far to the right when comparing the minor axis
surface brightness profiles in the first and second epochs.
Currently, we concern ourselves with the presence of the apparent
shift which, if confirmed, may be compared to the observed proper
motion of B2224+65, establishing a connection between the pulsar and
jet-like feature.  In order to determine the shift between epochs, we
apply a simple one dimensional edge detection method to the sharp
edges in the two epoch data.  Our edge detection method involves
constructing finely binned intensity profiles taken from high
resolution count maps (with $\sim$0.12$\arcsec$ pixels) and convolving
the profiles with a filter based on the work of \cite{canny86}, which
can be approximated by $\sim x$e$^{-\frac{x^2}{s^2}}$, with a filter
scale of $s$=0.5\arcsec, comparable to the ACIS resolution.  To
determine the underlying distribution of edge positions in the two
epoch data, and thus derive uncertainty estimates, we generate 5000
Poisson realizations of the high resolution count profiles and convolve
each with the chosen filter, following the method of bootstrap
resampling.  By filter design, the edge position for each realization
is determined by ``centroiding'' the peak in the surface brightness
profile and therefore depends primarily on the counting statistics of
the available data.  The derived positions are thus insensitive to the
exact choice of the filter scale or binning of the count profiles; we
have verified this using a variety of profile binning and filter
scales, all of which produce results consistent with those presented
here.  Based on the histogram distributions of the resampling
(Fig.~5), we find a shift between the two epochs of
0.62$^{+0.35}_{-0.28}$ arcsec, with 90 percent confidence
intervals. The reported errors on the shift are likely underestimates
of the true uncertainty as they do not account for errors in our
astrometric corrections, though they do include any possible
systematic uncertainty from unresolved bending in the jet.
Nevertheless, the shift, especially its direction, is consistent with
the proper motion of B2224+65, correcting for its projection to the
adopted minor axis of the jet (a factor of $\sim$0.88), and thus
supports an association between the pulsar and jet-like feature.

As mentioned previously, there is evidence for excess extended
emission along the minor axis of the jet in the second epoch compared
to the first.  This excess emission can also be seen in Fig.~2a,b
through the apparent widening of the jet-like feature over the two
epochs.  Examining the two epoch surface brightness profiles, there is
little deviation between epochs, after correcting for the previously
determined shift, except for the excess emission region, located
$\sim$7-17$\arcsec$ away from the left sharp edge.  The presence of
this region will not effect our previous edge determination due to its
distance from the left sharp edge and our choice of filter.
Following a method similar to the extension along the major
axis, the region of excess extended emission is found to have a
count rate that is 0.28$\pm$0.11$\rm~counts~s^{-1}$ higher in the
second epoch than in the first, which has a total count rate (sky
background included) of 0.64$\pm$0.06$\rm~counts~s^{-1}$.  Given the
high detection significance of the excess emission in the second epoch
($\sim$5$\sigma$), it is very unlikely that it is artificial or
statistical in nature; although a physical explanation for the
excess emission is perhaps even more enigmatic than the jet or
counter feature. 

\begin{deluxetable*}{llccccr}
\tablecaption{Spectral Fitting of the Jet-like Feature}
\tablehead{Epoch& Segment &CR (0.3-7.0 keV)&N$_H$ & $\Gamma$ & Flux
  (0.3-7.0 keV)&$\chi^2/d.o.f.$\\
& &10$^{-3}$ cts/s& 10$^{22}$ cm$^{-2}$&& 10$^{-14}$ ergs cm$^{-2}$ s$^{-1}$}
\startdata
\multicolumn{7}{c}{Two Epoch}\\
1&      Jet      &5.79$\pm$0.46& $<$0.12& 1.79$^{+0.41}_{-0.35}$& 5.16$^{+1.56}_{-1.62}$& 46.6/46\\
2&      Jet      &5.60$\pm$0.49& \nodata& 1.15$^{+0.43}_{-0.39}$& 7.38$^{+1.99}_{-2.18}$&\nodata\\
Merged& B2224+65 &1.86$\pm$0.14& \nodata& 1.88$^{+0.44}_{-0.37}$& 1.68$^{+0.30}_{-0.44}$&\nodata\\
\multicolumn{7}{c}{Epoch Averaged}\\
Merged& Jet      &5.70$\pm$0.34& $<$0.15& 1.61$^{+0.35}_{-0.17}$& 5.58$^{+0.77}_{-1.22}$& 52.7/48\\
Merged& B2224+65 &1.86$\pm$0.14& \nodata& 1.95$^{+0.44}_{-0.21}$& 1.63$^{+0.28}_{-0.52}$&\nodata\\
\multicolumn{7}{c}{Major Axis}\\
Merged&  Head Segment& 3.61$\pm$0.25& $<$0.13& 1.69$^{+0.39}_{-0.34}$& 3.59$^{+0.55}_{-0.84}$ & 33.6/47\\
\nodata& Tail Segment& 1.96$\pm$0.21& \nodata& 1.00$^{+0.53}_{-0.47}$& 3.35$^{+0.80}_{-1.13}$ &\nodata\\
\nodata& B2224+65&     1.86$\pm$0.14& \nodata& 1.90$^{+0.43}_{-0.37}$& 1.65$^{+0.31}_{-0.40}$ &\nodata\\
\multicolumn{7}{c}{Minor Axis}\\
Merged&  Edge Segment&  2.64$\pm$0.22& $<$0.06& 1.08$^{+0.43}_{-0.32}$& 3.76$^{+0.78}_{-0.57}$ & 31.9/45\\
\nodata& Taper Segment& 2.79$\pm$0.24& \nodata& 1.30$^{+0.51}_{-0.37}$& 3.70$^{+0.98}_{-0.76}$ & \nodata\\
\nodata& B2224+65&      1.86$\pm$0.14& \nodata& 1.73$^{+0.42}_{-0.27}$& 1.77$^{+0.31}_{-0.30}$ & \nodata
\enddata
\tablecomments{\textsc{xspec} spectral fitting parameters for the jet
  similar to Table 2. Column densities are again given as upper
  limits.  The inclusion of B2224+65 with the jet segments serves to
  jointly constrain N$_{\rm{H}}$ and reduce the errors associated with the
  relevant $\Gamma$s.  The epoch averaged jet spectrum is determined
  using the same method as the epoch averaged pulsar spectrum.  Errors
  are given to 90 percent confidence.}
\end{deluxetable*}

\begin{figure}
\includegraphics[width=.5\textwidth]{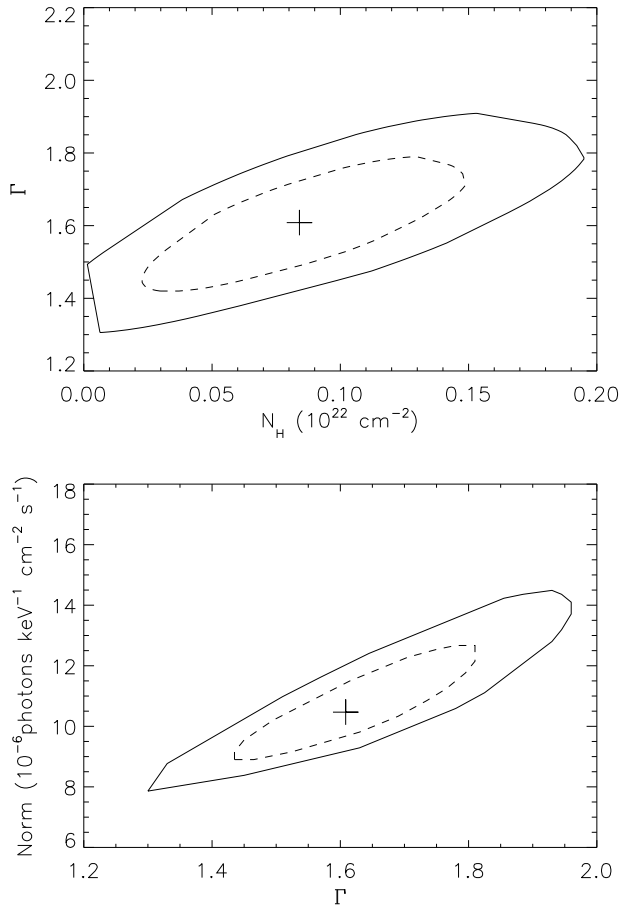}
\caption{68 percent (dashed line) and 90 percent (solid line)
  confidence contours for the merged jet spectral fit similar to Fig.~3.}
\end{figure}

We now examine the spectrum of the jet both spatially and across
epochs to determine if there may be evidence for any spectral
evolution.  In each epoch, we extract the spectrum of the jet
using the same rectangular region (see Fig.~1,2) after removing B2224+65.  The
jet region is further segmented along its major and minor axes,
labeled according to their relative positions, such that the segments
have approximately equal counting statistics.  The individual spectra
are again adaptively binned to a S/N ratio of
$\ge$2.5, resulting in $\gtrsim$25 counts per bin due to the higher
background contribution in the jet region.  In order to improve the
constraints on model parameters in the spectral fits for the segments, we
again apply joint fits in which the parameters are set to be the
same in both epochs though allowed to vary over the individual
segments; only N$_{\rm{H}}$ is set as a common parameter throughout,
assuming that it is constant in time and space across the jet.
We also include the epoch averaged pulsar spectrum to place further
constraints on N$_{\rm{H}}$ and the relevant $\Gamma$s.  The fitted
parameters are given in Table 3 with the confidence contours for the
epoch averaged jet spectrum given in Fig.~6.  Similar to
\cite{hui07}, we find that the power law model provides a satisfactory
fit ($\chi^2/\rm{d.o.f}\sim$1.1) to the jet spectrum, though the
possibility of alternative models can not be completely rejected based
on $\chi^2$ test alone.  It should be noted, however, that reasonable
alternatives such as thermal Bremsstrahlung result in temperatures on
the order of $\ga$2 keV \citep[see also][]{hui07} and are likely
unrealistic. From our spectral analysis, there is evidence for
significant, to 95 percent confidence, spectral hardening in
the jet between epochs.  There is also indication for further
spectral variation along the jet, including spectral hardening
farther from the pulsar and softening counter to the pulsar proper
motion, although these trends are only marginally detected (with
$\sim1\sigma$ significance).

\section{Discussion}

Based on our two epoch X-ray analysis of B2224+65 and the jet-like
feature, we have verified the proper
motion of the pulsar and have shown similar motion in the sharp
leading edge of the jet, providing strong evidence for its association
with the pulsar.  Along the major axis of the jet, a region of extended X-ray
emission (hereafter counter feature), detected with $\sim$4$\sigma$
confidence, is found near the pulsar and opposite the jet.  The
jet also appears wider, due to a significant ($\sim$5$\sigma$) region of
excess emission, with a harder spectrum in the second
epoch compared to the first whereas additional spectral evolution across the
major and minor axes is more ambiguous.  Given the low confidence for
many of the spectral variations, including softening counter to the
direction proper motion, the physical nature of the jet-like feature remains
puzzling.  Here, we briefly examine two candidate physical scenarios
for the production of the jet based on current results.  

To begin with, let us first consider the model proposed by
\cite{band08}. This model assumes that the jet represents electrons
that have been accelerated at the termination shock of the pulsar wind
which then leak out of the bow shock and diffuse into an ambient
medium containing a large-scale plane-parallel magnetic field. The
orientation of the jet-like feature feature would then correspond to
the direction of the magnetic field, which needs to be unusually strong ($\sim
45\mu$G). As stated by \cite{band08}, this scenario would predict
motion for the jet-like feature consistent with B2224+65, softening of the
jet spectrum counter to the direction of proper motion due to
synchrotron cooling of particles injected at different times and the
presence of a counter feature.  Our two epoch observations support
many of these predictions though the spectral variation
can not be positively confirmed.  Additional spectral variations, such
as hardening between epochs and away from the pulsar along the jet
major axis, are difficult to explain under this scenario.  If the
jet spectrum was observed to soften away from the pulsar, then one
may naturally expect synchrotron cooling to be responsible; however, the
observed hardening implies acceleration within
the jet feature, unlikely for particles confined by a magnetic field.
Hardening between epochs may suggest variation in the acceleration
of the energetic particles, possibly due to changes in the pulsar wind and/or
ISM properties.  However, the lack of spectral evolution for the
pulsar advocates against any such deviations at the terminal shock
if, in fact, a significant portion of the pulsar emission results
from the ISM-pulsar wind interaction.  Furthermore, the relative
faintness of the counter feature posses a challenge for the
\cite{band08} model, as the energetic electrons should have no
preference in which direction along the magnetic field they travel,
though this issue may be rectified by internal anisotropies and
asymmetry of the pulsar wind. 

Alternatively, the B2224+65 extended X-ray features may represent the
externally processed particles from jets, akin to those associated
with AGNs.  In this scenario, the main jet-like feature, counter
feature and the Guitar Nebula represent the pulsar jets and equatorial
wind outflow where the latter is responsible for the formation of the
Guitar Nebula trailing the pulsar.  X-ray emission is not expected in
the pulsar wind forming the Nebula as the particles are not energetic enough to
generate detectable X-ray synchrotron radiation.  The pulsar jets,
intrinsically relativistic and containing highly energetic particles,
may easily escape the bow shock but are ultimately confined by the
large ram-pressure generated by the pulsar's motion.  In this
scenario, the spectrum of the jet-like feature would soften counter to the
direction of proper motion, similar to the \cite{band08} model, due to
synchrotron cooling.  If enhancement of the X-ray emission due to the
high ram-pressure contributes significantly to the X-ray emission,
then spectral variation between epochs may arise from density
variations in the ISM resulting in post-processing variations of the
energetic particles (i.e. through their acceleration and/or
synchrotron cooling efficiency).  ISM variation over time has been
seen previously by \cite{chat04} where they found a decrease in the
local density by a factor of $\sim$0.7 between 1994 and 2001; however,
any such density contrast can not be directly confirmed given the current two
epoch \textit{Chandra} observations and the relatively static pulsar
spectrum.  Spatial variations of the ISM could also explain the spectral 
hardening away from the pulsar through similar post-processing of the
energetic particles.  The relative faintness of the
counter feature compared to the main jet-like feature may be due to
Doppler effects if they are indeed relativistic and aligned along
the line of sight; specifically, if $\beta\cos\theta\sim0.4$ where
$\beta$ is the bulk flow velocity and $\theta$ is the angle along the
line of sight. Further constraints may be possible by invoking more
detailed jet mechanics and synchrotron theory.

The ballistic jet scenario was deemed unlikely by \cite{band08}, as
it would predict a ratio between the length and width of the
jet-like feature ($L_{jet}$ and $W_{jet}$, respectively) that is too small to
match observations ($L_{jet}\sim0.6$ pc and $W_{jet}\lesssim0.04$ pc for
D$_{\rm{kpc}}=1$) along with significant bending in the jet based
on comparisons of the bow shock pressure, transverse momentum added to
the jet through the ISM ram-pressure and its the initial momentum.
Following \cite{band08}, we may estimate the expected bending
angle of the jet-like feature under the ballistic jet model by
assuming that the pulsar imparts a constant fraction $\eta$ of its
spin-down energy into the jet, which has a transverse size of
$W_{jet}$.  Balancing the transverse and initial momenta, the angle
$\theta$ describing the bending of the jet may be given by 
\begin{equation}
\tan\theta=\frac{W_{jet}L_{jet}\sin(118\degr)\delta \mu m_{H} n_{ISM}v^2_*}{\eta\dot{E}/c}
\end{equation}
where $\delta$ and $\mu$ are the ISM ionization fraction and mean
molecular weight, respectively, and $v_*$ is the pulsar proper
motion.  If we assume a fully ionized ISM composed mostly of hydrogen
($\delta=1$ and $\mu=0.5$) with number density $n_{ISM}\approx
0.1\rm~cm^{-3}$ and dimensions for the proposed jet as taken from current
observations, then such a ballistic jet should trail behind the pulsar
regardless of the value of $\eta$; strongly suggesting against the jet
interpretation.  However, the actual jet may be very narrow
($W_{jet}\ll 0.01$ pc), similar to AGN jets, and/or the ISM could be
largely neutral in which case the expected bending angle would be
significantly reduced. A largely neutral ISM would contribute little to the
ram-pressure as the mean free path of neutral atoms is much larger
than the width of the feature ($\sim$0.1 pc for neutral hydrogen atoms
with density $0.1\rm~cm^{-3}$); however, this would impose
a fine-tuning problem on the jet-ISM interaction where the number
density for the small ionized fraction must still be large enough to
generate the confining ram-pressure and accelerate the energetic
particles to X-ray emitting energies.  An additional complication of
the jet interpretation is the misalignment of the spin axis (believed
to be correlated with pulsar jets) and the direction of B2224+65's
proper motion; nevertheless, the method by which pulsars or their
progenitors acquire such high velocities remains uncertain \citep[see, for
  example,][]{hills88,yu03}.  Certainly, additional and more detailed
theoretical modeling, which may require additional observations in
order to confirm apparent spectral trends, will be necessary to
further test the jet and \cite{band08} models.

\section{Summary}

Using the available two epoch \textit{Chandra} observations, we have
presented an X-ray morphological and spectral study of the pulsar
B2224+65 and its apparent extended linear X-ray-emitting features.
We confirm B2224+65's proper motion over the two epochs and find
little variation in its spectrum.  Additionally, we find motion in the
sharp leading edge of the jet-like feature that is consistent with the
proper motion of the pulsar.  The jet is found to be
significantly harder and wider in the second epoch than the first with
spectral variation along its major and minor axis detected only
marginally. These observations are somewhat consistent with
magnetically confined ``leaking'' of energetic particles as proposed
by \cite{band08} as well as a typical jet interpretation; although
there are certain limitations to each of these scenarios.  In the \cite{band08}
model, it is difficult to explain the faintness of the counter
feature as well as the spectral variations between epochs, if verified.  On the
other hand, it is difficult to imagine a jet being responsible for the observed
emission unless the jet is actually very thin and unresolved, similar
to AGN jet outflows, or if the ISM is mostly neutral, which would then
impose fine-tuning of the ISM number density and ionization fraction.
Detailed spectral and theoretical modeling following pulsar ejecta
interacting with the ISM and/or an ambient magnetic field, including
turbulent effects within the ejecta, will be necessary in order to
place any further constraints on the physical origin of the jet
given current observations.  Additional observations may then be
required for confirmation of the observed spectral trends and place
definite constraints on the origin of this enigmatic X-ray feature.

\acknowledgments

We would like to thank the anonymous referees for their helpful
comments on improving the manuscript and acknowledge Eric Gotthelf for his
comments in developing the calibration and analysis for the extended
feature. This work is supported by NASA under the grant NNX07AU28G. 


\end{document}